\documentclass[12pt]{article}
\usepackage{ulem}
\usepackage{times}
\usepackage{color}
\usepackage{graphicx}
\usepackage{calc}
\usepackage{epsfig}
\usepackage{float}
\usepackage{amsfonts}

\begin{document}

\parskip 18pt
\baselineskip 0.25in

\begin{center}
{\LARGE \bf Bursts generate a non-reducible spike pattern code}
\end{center}
\vspace{0.5cm}

\begin{center}
{\bf  Hugo G. Eyherabide$^{1,2}$, Ariel Rokem$^{1}$, Andreas V. M.
Herz$^{1,3}$, and In\'es Samengo$^{2*}$}
\end{center}

\begin{enumerate}
\item[$^1$] Institute for Theoretical Biology, Department of
Biology, Humboldt Universit\"at, and Bernstein Center for
Computational Neuroscience Berlin, Berlin, Germany.

\item[$^2$] Centro At\'omico Bariloche and Instituto Balseiro,
San Carlos de Bariloche, Argentina.

\item[$^3$] Department of Biology,
Ludwig-Maximilians-Universit\"at, and Bernstein Center for
Computational Neuroscience Munich, Martinsried, Germany.

\end{enumerate}

\noindent {\bf Correspondence}
\\In\'es Samengo\\ Centro At\'omico Bariloche\\
San Carlos de Bariloche, (8400), R\'{\i}o Negro, Argentina.\\ Tel:
++ 54 2944 445100 (int: 5391 / 5345). Fax: ++54 2944 445299.\\
Email: samengo@cab.cnea.gov.ar

\noindent {\bf Running title:}\\ Burst-mediated neural coding.

\pagebreak

\section*{Biography}

\begin{figure}[htdf]
\includegraphics{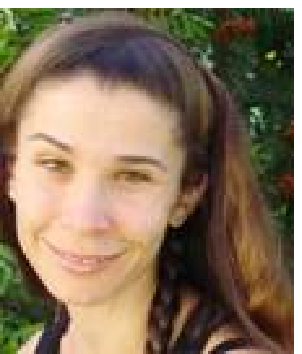}
\end{figure}

\noindent In\'es Samengo studied physics in Argentina, and then
switched to neuroscience. She spent a year in Berlin, in the group
of Andreas Herz, where she also met Ariel Rokem. Together they
studied the neural code of insect sensory systems. Back in
Argentina, In\'es settled as a researcher in Bariloche, where Hugo
Eyherabide joined her as a PhD student. Hugo and In\'es still
interact regularly with Andreas (now in Munich) and Ariel (now in
Berkeley), resulting in a quite entertaining inter-continental
collaboration group.

\section*{Abstract}

At the single-neuron level, precisely timed spikes can either constitute firing-rate codes or
spike.pattern codes that utilize the relative timing between consecutive spikes. There has been
little experimental support for the hypothesis that such temporal patterns contribute substantially to information transmission. By using grasshopper auditory receptors as a model system, we
show that correlations between spikes can be used to represent behaviorally relevant stimuli.
The correlations reflect the inner structure of the spike train: a succession of burst-like patterns. We demonstrate that bursts with different spike counts encode different stimulus features, such that about {\it 20\%} of the transmitted information corresponds to discriminating between different features, and the remaining {\it 80\%} is used to allocate these features in time. In this spike-pattern code, the what and the when of the stimuli are encoded in the duration of each burst and the time of burst onset, respectively. Given the ubiquity of burst firing, we expect similar findings also for other neural systems.

\section*{Keywords}

\noindent Burst spiking, neural code, sensory encoding,
information theory, auditory receptor.

\pagebreak


\section{Neural codes based on spike-timing}

Single neurons can encode time-dependent stimuli in several ways. In {\bf firing-rate codes} the only
response feature that carries information is the time-varying
firing probability. The information encoded in subsequent spikes (if any) is redundant with the one available from the firing rate. The firing probability is a
function of the stimulus, often non-local and non-linear.
Depending on the nature of the transformation between the stimulus
and the firing rate, the activity of the neuron may encode the
stimulus continuously, or may instead only extract one or a few
specific features.

\begin{figure}[htdf]
\includegraphics{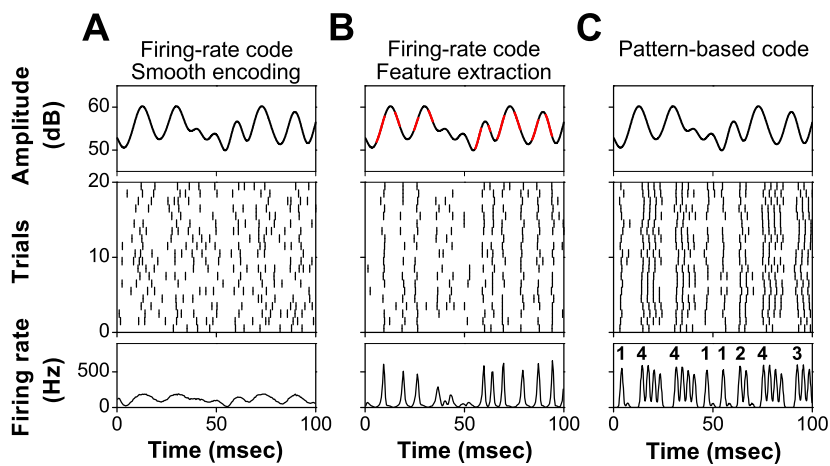}
\caption{\label{f1} {\bf Comparison between different neural
codes.} A: Firing-rate code, where the strength of the stimulus is
encoded in the number of spikes fired in an extended time window.
The temporal evolution of the firing probability (shown below)
mimics the stimulus. B: Firing-rate code, where the occurrence of
specific stimulus features (in this case, pronounced up- or
downstrokes marked in red) is encoded in the times at which
individual spikes are generated. (C) Pattern-based codes, in which
different stimulus features are represented by different spike
patterns. In this case, a burst-mediated code is shown, where
patterns are distinguished by their intra-burst spike count.}
\end{figure}

These two extreme situations are depicted in the first two panels
of Fig.~\ref{f1} A,B. In Fig.~\ref{f1}A, the firing probability is larger than zero almost
everywhere, and the cell fires more than one spike in
intervals comparable to those of the fluctuations in the stimulus.
The number of spikes per unit time encodes the stimulus strength
through a monotonic transformation. By reading out the number
of spikes in an extended time window, typically in the order of
tens of milliseconds, downstream neurons have access to a smooth
representation of the stimulus.

Fig.~\ref{f1} (B) also exemplifies a firing-rate code, but in this case the firing probability does not represent the entire dynamic range of the stimulus. It rather picks specific features, so that spike generation is only possible shortly after these particular features. In these codes, each single spike suffices to inform downstream neurons of the presence of the relevant feature, and the temporal jitter of individual spikes is often smaller than 1 millisecond. 

The two examples mentioned above fall within the broad class of firing-rate codes. In one extreme, we find codes where the firing rate varies slowly, so the precise timing of individual spikes is not crucial. The neuron encodes the stimulus strength making use of a fairly broad dynamical range, but does so with low temporal precision. In the opposite extreme, the firing rate varies rapidly. As the cell only represents either the presence or absence of the relevant feature, little information is provided about the overall evolution of the stimulus. However, the rapid variation of the time-dependent firing rate provides large amounts of information about the temporal location of the encoded features. These two extreme cases can be framed in a unified mathematical formulation (Rieke et al., 1997). Depending on whether the relevant feature is sharp and brief or flat and broad, the system ranges from a firing-rate code based on individual spikes to a firing-rate code based on mean spike counts.

Starting with the seminal work of Mainen and Sejnowski (1995), several studies have shown that precise spike timing down to the sub-millisecond regime is important to transmit information about the sensory world. Examples include the insect visual system (Strong et al., 1998), the vertebrate lateral geniculate nucleus (Reinagel and Reid, 2000), the rodent somatosensory thalamus (Montemurro et al., 2007a), and the auditory system, e.g., invertebrate receptor cells (Rokem et al., 2006), vertebrate brainstem neurons (Oertle, 1999) and auditory cortical cells (Heil, 1997).

However, spike-timing based codes can go beyond firing-rate codes. In principle, not only the location of precisely timed individual spikes can transmit information, but also the relative timing between two or more spikes. Such schemes are referred to as relational, or {\bf spike-pattern codes}. In such codes, the correlations between spikes define patterns, and these patterns are employed to encode stimulus features, each feature corresponding to a particular sequence of {\bf inter-spike intervals} ({\bf ISIs}). Reich et al. (2000) and Doiron et al. (2007) showed that different patterns (in this case doublets of spikes separated by ISIs of different durations) were associated with different stimulus features. In these examples, however, different patterns have different instantaneous firing rates, so unless some additional characterization is made, one cannot rule out a purely firing-rate coding scheme.

A neural code that cannot be explained in terms of the instantaneous firing rate requires more complex patterns, including three or more spikes (two or more ISIs). In this paper we review the coding capabilities of a ubiquitous burst-mediated code, where the distinction between different patterns is given by the number of spikes that compose each {\bf burst}. The set of different codewords consists of bursts containing different numbers of spikes, as exemplified in Fig.~\ref{f1} (C). In this case, the intra-burst ISI for different codewords remains approximately the same, and the distinction between the different code-words is given by the intra-burst spike count.

Traditionally, burst firing was believed to underlie unconscious regulatory processes during sleep, seizures or anesthesia, and to prevent sensory signals from reaching higher processing stages. This picture emerged from the robust synchronized bursting activity that rises spontaneously in thalamic slices, even in the absence of stimulation (Guillery, 2001). Bursting is also observed in thalamic neurons during sleep, and is disrupted as soon as the subject wakes up, to be replaced by tonic activity during wakefulness. However, in the last decade several studies have shown that burst firing also participates in the representation of the sensory world during the aroused state (Sherman, 2001), as well as in other neural systems (Krahe and Gabiani, 2004). For example, in the electrosensory lateral line lobe of the weak electric fish, bursts represent low-frequency events (Oswald et al., 2004), comprising either excitatory or inhibitory stimulus deflections (Metzner et al., 1998). In the mammalian LGN, bursts encode slow stimulus features (Lesica and Stanley, 2004), characterized by high contrast (Reinagel et al., 1999), typical of natural images (Denning and Reinagel, 2005). In the rodent hippocampus, bursting pyramidal place cells represent the location of the animal in the environment, both through the firing rate (Wilson and McNaughton, 2003) and the timing with respect to the theta cycle (O'Keefe and Reece, 1993). 

A few studies have specifically explored the information carried by the intra-burst spike count. Two of them involve the mammalian primary visual cortex (DeBusk et al., 1997; Martinez-Conde et al., 2002), where the length of each burst was correlated with the orientation of the stimulus. A theoretical analysis based on a computational model of a cortical pyramidal cell (Kepecs and Lisman, 2003) concluded that the number of spikes inside each burst represented the slope of the incoming stimulus at burst onset. Finally, in tactile sensory neurons in leech (Arganda et al., 2007), the intra-burst spike count represented the velocity of skin displacements. These analyses demonstrate that bursting neurons in different systems represent different stimulus attributes. A common aspect however, is that the information is not only encoded in the time-dependent firing rate, but also in the correlations between spikes. 

\section{Burst-mediated codes in grasshopper auditory receptors}

Grasshoppers communicate with each other by chirping acoustic signals produced by rasping their hind legs across their wings. By analyzing the response of acoustic receptor cells to a broad range of naturalistic and artificial stimuli, we have demonstrated that spiking activity is particularly precise when driven by sound waves whose temporal characteristics coincide with those of the natural songs (Rokem et al., 2006). When these stimuli are played at moderate or loud volume (e.g., nearby sources), receptor neurons have a large probability to elicit bursting responses (Eyherabide et al., 2008). Actually, 93\% of the recorded neurons generated bursts when driven with naturalistic stimuli, whereas none of them bursted in response to signals that varied much faster than the natural songs. Hence, bursting seems to appear in response to behaviorally relevant stimuli only.

\begin{figure}
\includegraphics{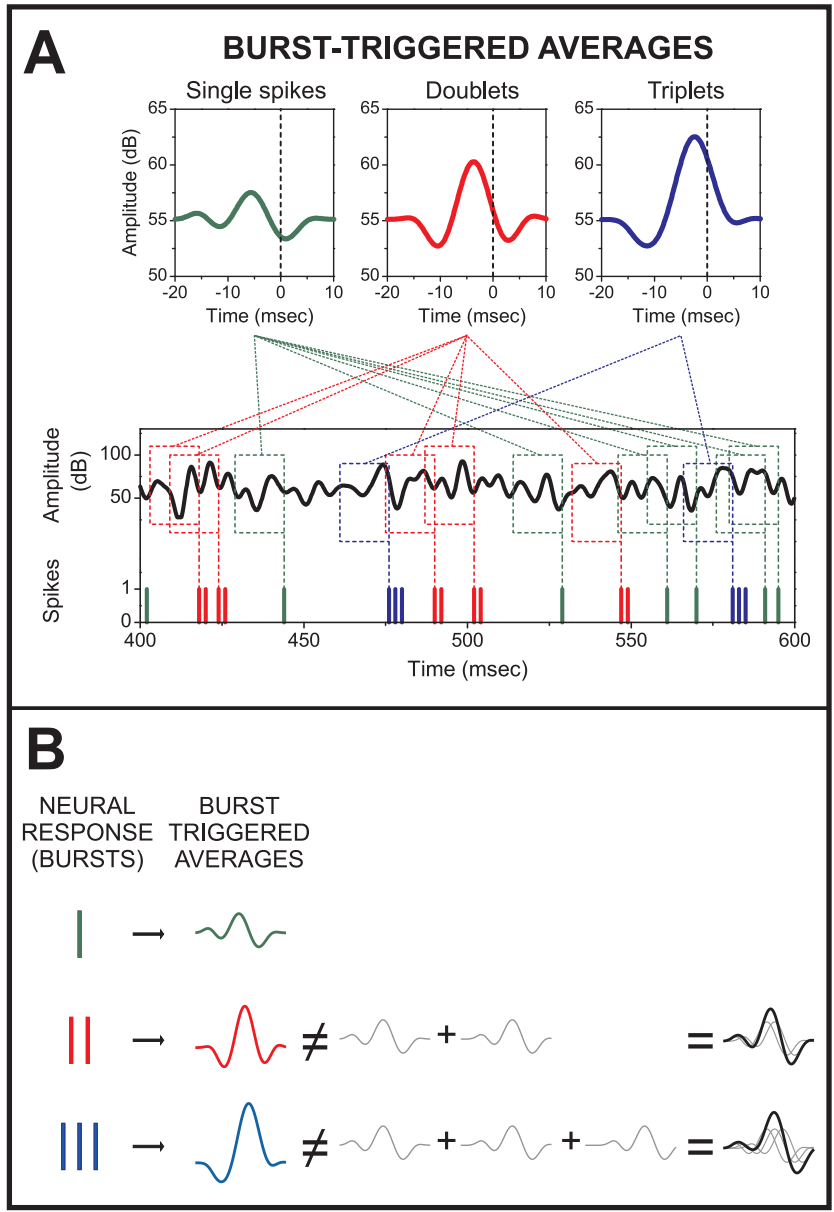}
\caption{\label{f2} {\bf The burst-triggered averages (BTAs)
corresponding to bursts of different lengths.} A: Calculation of
BTAs. The stimulus stretches preceding bursts of exactly $n$
spikes are averaged together, and compared to one another. For all
cells, BTAs exhibit a central peak, on top of a noisy background.
The size of the peak increases with the number of spikes in the
burst. B: The BTA corresponding to doublets is larger than the one
that would be obtained by placing two copies of the BTA of single
spikes, displaced from one another by one ISI. The same holds for
triplets and quadruplets.}
\end{figure}

If bursts of different spike counts are mapped onto different stimulus features, then the stimuli eliciting shorter bursts must be significantly distinct from the stimuli generating longer bursts. The easiest way to test this hypothesis is to compare the average stimulus preceding single spikes, with that corresponding to doublets, triplets and quadruplets. We defined the {\bf burst triggered average (BTA)} corresponding to bursts of $n$ spikes as the average stimulus time course before bursts containing exactly n action potentials (Fig. ~\ref{f2}(A)). 

In our case, the time-dependent stimulus was the volume (in decibels) of the envelope of a high-frequency input sound wave (Eyherabide et al., 2008). All the BTAs exhibited a pronounced peak on top of a noisy background. Hence, burst production occurred a few milliseconds after a sudden elevation of the sound intensity. The magnitude of the elevation determined the number of spikes in each burst, such that higher stimulus fluctuations elicited longer bursts. In 85\% of all bursting cells, this correspondence was selective: BTAs associated to single spikes, doublets, triplets and quadruplets were significantly different from each other. As a consequence, the number of spikes per burst was a good predictor of the maximal height of the transient intensity fluctuation (Eyherabide et al., 2008). The relationship between stimulus intensity and burst duration was not trivial, as sketched in Fig. 2 (B). In 95\% of the bursting neurons, the mean stimulus eliciting a doublet was significantly larger than the stimulus that would be obtained by summing up two copies of the mean stimulus generating isolated spikes, separated by the observed ISI. 

\begin{figure}
\includegraphics{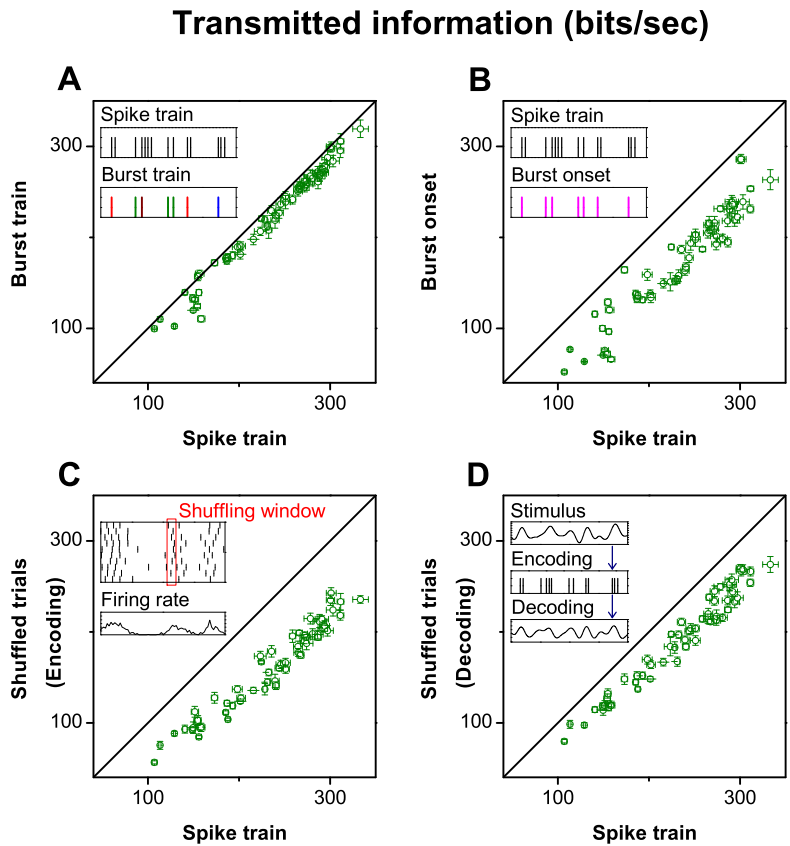}
\caption{\label{f3} {\bf Comparison between the information rate
carried by manipulated spike trains (in the $y$-axis) and the
information rate in the original response ($x$-axis), for all
bursting cells in the population}. In all cases, information rates
were calculated using the direct method (Strong et al., 1998), with
the error estimations from Montemurro et al., (2007b). A: The spike
train is parsed into a sequence of bursts that are only
distinguished by their spike count, such that their inner structure
is neglected. Inset: Green, red, blue and brown symbols in the burst
train represent bursts containing 1, 2, 3 or 4 action potentials in
the original spike train, respectively. The sequence of bursts
contains 94\% of the information of the full spike train. B: If the
same symbol is used to represent bursts of different lengths, then
distinctions between different stimulus features is no longer
possible (see inset). Only temporal information referring to the
timing of each feature remains. This implies a 22\% reduction of the
original information (21\%, if compared to the information
transmitted by different types of bursts). C: If trials within each
fixed time bin are shuffled, all within-trial temporal correlations
are destroyed, though the time-dependent firing rate is preserved.
As a consequence, the information drops by 32\%. D: If the stimulus
is decoded from the spike train (see inset) without taking temporal
correlations into account, the information is reduced in 18\%.}
\end{figure}

Differentiating between bursts of different durations, hence, allows us to distinguish between different stimulus features. But how can we be sure that there is no other alphabet that could do a better job? Perhaps there is another set of patterns that also allows one to differentiate between different stimulus features. We would like to have a criterion to quantify the adequacy of the chosen alphabet, and to rank it with respect to other possible choices. One way would be to measure how much information is lost by only distinguishing between the patterns of the chosen alphabet, and to neglect all additional response features. For a burst code, this amounts to distinguishing between bursts containing different number of spikes, while disregarding the internal temporal structure of each burst beyond its spike count. Not all doublets have exactly the same ISI between their two spikes, and not all triplets have exactly the same ISI sequence.  In order to assess the success of the burst alphabet we therefore neglect these differences. Operationally, the spike train is represented as a sequence of symbols that specify the number of spikes in each burst, as sketched in the inset of Figure ~\ref{f3} (A). All doublets are represented by the same symbol, irrespective of their inner structure. The same holds for triplets, quadruplets, and higher-order bursts. The resulting sequence of symbols can be used to calculate the mutual information between the stimulus and the burst alphabet. If this information is substantially lower than the information in the original spike train, then the inner structure of bursts must be considered as relevant, and the selected alphabet as not appropriate. But this is not the case with our data. In Figure ~\ref{f3} (A) we show that the burst representation has almost the same amount of information as the full response. This implies that bursts encode different stimulus features essentially through their intra-burst spike count.   

The burst code, though missing several response details, still provides an informative and compact representation of the original spike train. The advantage of such a representation is that now we can explicitly interpret the code. At this stage, we know that the number of spikes in each burst represents the height of the stimulus feature that elicited bursting, whereas the time at which the burst is generated tags the temporal location of the relevant feature. One may therefore wonder how much information corresponds to distinguishing between different features, and how much is accounted for allocating them in time. To show this, we compared the information in the burst train with that of an even more drastically reduced representation of the spike train, in which all bursts are mapped onto the same symbol. No distinctions between different types of bursts (and stimulus features) thus remain, as shown in Fig. 3 (B). In the investigated cells this leads to an average reduction of the encoded information of 22\%. This percentage represents the 'what' in the stimulus, whereas the complementary fraction accounts for the 'when' (Berry et al., 1997;  Theunissen and Miller, 1995; Borst and Theunissen, 1999).   

\section{Is the burst code a firing rate code?}

In our data, the instantaneous firing rate inside different bursts is always approximately the same: we found no significant dependence of the intra-burst ISI on the number of spikes per burst. Yet, it could be argued that the burst code is still a firing-rate code, but such that the mean firing rate of the cell should be read out in long intervals. In Rokem et al. (2006) we showed that the larger the time bin used to read out neural responses, the smaller the amount of transmitted information. The loss arises because longer bins have less temporal precision.  One could then try using broad time bins, but make them slide along the spike train in very fine steps. Even so, the number of spikes per burst cannot be obtained by counting the spikes inside a window of fixed duration. The code is composed of sequences that alternate between short and long bursts, and the interval between bursts is often comparable to the duration of bursts themselves. Hence, if the window used to count spikes is short, long bursts are not captured. Instead, if a long window is used, then several short bursts are mistaken as a single longer burst.  The same problem arises if one tries to convolve the spike train with a smooth bell-shaped weight function, as for example, a Gaussian kernel of fixed width. Therefore, bursts do not constitute a convolved firing-rate code either.   

More importantly, the burst code found in grasshopper receptors uses not only the precision of individual spikes, but also the correlations between spikes. Two subsequent spikes may or may not be part of the same burst, depending on the size of the ISI separating them. Hence, it is their relative timing that matters. In firing rate codes, however, correlations between spikes can be entirely explained in terms of the time-varying firing probability. Therefore, correlations make no additional contribution to the encoding or decoding of information, beyond the information available in the firing probability.   

The impact of correlations on the encoding of sensory information can be assessed by shuffling trials in the neural response, for each fixed time bin, as highlighted in the inset of Figure ~\ref{f3} (C). By doing so, the time-dependent firing rate is preserved, while the within-trial correlations are abolished. If the shuffled spike train contains significantly less information than the real spike train, then removing the correlations has a negative impact in the transmitted information.  This approach was proposed as a measure of conditional independence in the framework of population coding (Schneidman et al., 2003). When later employed to assess the role of temporal correlations in single neurons in the rat thalamus (Montemurro et al., 2007a), spike patterns played a minor role. They only increased the total information by 6\% and the remaining 94\% was entirely attributable to the time-dependent firing rate. Thus, most information was transmitted by precisely timed single spikes. In grasshopper receptor cells, however, spike patterns play a more important role, as shown in Figure ~\ref{f3} (C). The shuffled responses carried 32\% less information than the full spike train. The constraints imposed by correlations, hence, resulted in a substantially improved coding scheme as compared to the one that would be obtained by independent time bins.   

A complementary approach is to assess the effect of correlations in neuronal decoding (Latham and Nirenberg, 2005). This formulation allows us to evaluate whether the original stimulus can be decoded from the neural responses equally well, if within-trial temporal correlations are neglected. Since we do not know which decoding scheme is employed by downstream neurons in the brain, we use optimal Bayesian decoding, guaranteed to perform at least as well as any biological decoding. Figure ~\ref{f3} (D) shows that neglecting the correlations has a detrimental effect in the decoded information of almost 20\%, an amount that is very similar to the information needed to discriminate between different stimulus features, see panel (B). Therefore, grasshopper auditory responses can only be decoded properly if their correlational structure is taken into account as summarized in Figure 4.   

\begin{figure}[htdf]
\includegraphics{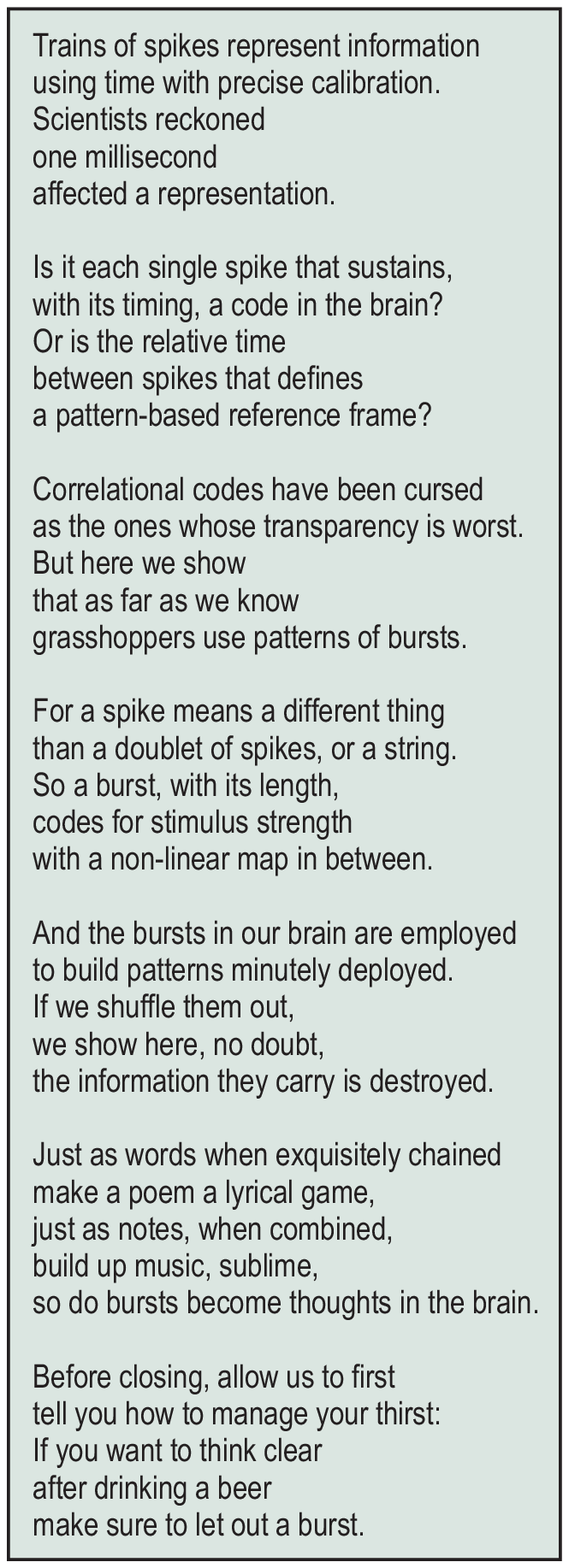}
\caption{\label{f4} {\bf A burst of rhymes to close, and complement the prose.}}
\end{figure}

\section*{Conclusions}

One of the central problems in neuroscience is to understand the way in which sensory information is represented in the nervous system. The objective is to know the general principles of the encoding scheme, and to have an explicit dictionary connecting stimulus features with response characteristics. This study addresses both issues. By using information-theoretical quantities, we were able to extract some general principles governing the representation of acoustic information in grasshopper receptors. We demonstrate that the code is based on spike patterns, and that the information conveyed by these patterns cannot be accounted for by a rapidly varying firing rate. The code is structured in bursts, and the distinctive feature discerning between different types of bursts is the number of spikes they contain. The time of burst onset represents the temporal location of a certain stimulus feature, and the intra-burst spike count discriminates between different types of features. Using theoretical techniques, we quantified the fraction of information that corresponds to these two aspects individually. By thoroughly examining the correspondence between bursts and stimuli, we revealed the meaning of each pattern in terms of the height of sudden amplitude fluctuations. The non-linear transformation between spike count and stimulus amplitude provides additional evidence that spike patterns must be read out as compound codewords, and not as a collection of single spikes.   

Our model system, hence, provides an example of a correlation-based neural code whose building blocks are bursts of spikes. Burst firing is ubiquitous in the nervous systems of both vertebrates and invertebrates. Our study suggests that the role of bursts in sensory representations might be even more relevant than previously thought.   

\section*{Acknowldgements}

This work was supported by the Alexander von Humboldt Foundation,
the Deutsche Forschungsgemeinschaft (SFB 618), the Consejo de
Investigaciones Cient\'{\i}ficas y T\'ecnicas, the German Federal
Ministry of Education and Research, the Israeli Ministry of
Science, the Minerva Foundation of the Max Planck Society, and the
Agencia de Promoci\'on Cient\'{\i}fica y Tecnol\'ogica of
Argentina.

\section*{References}

\begin{enumerate}

\item[-] Arganda S., Guantes R., and de Polavieja G. G. (2007).
Sodium pumps adapt spike bursting to stimulus statistics. Nat.
Neurosci. 10, 1467-1473.

\item[-] Berry M. J., Warland, D. K., and Meister, M. (1997). The
structure and precision of retinal spike trains. Proc. Natl. Acad.
Sci. USA 94, 5411-5416.

\item[-] Borst, A., and Theunissen, F. E. (1999). Information
theory and neural coding. Nat. Neurosci. 2, 947–957.

\item[-] Denning K. S., and Reinagel P. (2005). Visual Control of
Burst Priming in the Anesthetized Lateral Geniculate Nucleus. J.
Neurosci. 25, 3531–3538.

\item[-] DeBusk B. C., DeBruyn E. J., Snider R. K., Kabara J. F.,
Bonds A. B. (1997). Stimulus-Dependent Modulation of Spike Burst
Length in Cat Striate Cortical Cells. J. Neurophysiol. 78,
199-213.

\item[-] Doiron B., Oswald A. M. M., and Maler, L. (2007).
Interval Coding. II. Dendrite-Dependent Mechanisms. J.
Neurophysiol. 97, 2744–-2757.

\item[-] Eyherabide, H. G., Rokem, A, Herz A. V. M, and Samengo,
I. (2008). Burst firing is a neural code in an insect auditory
system. Front. Comput. Neurosci. 2:3. \\
doi:10.3389/neuro.10.003.2008

\item[-] Guillery, W. (2001). Exploring the thalamus. (New York:
Academic Press).

\item[-] Heil, P. (1997). Auditory cortical onset responses
revisited. I. First-spike timing. J. Neurophysiol. 77, 2616--2641.

\item[-] Kepecs A., and Lisman J. (2003). Information encoding and
computation with spikes and bursts. Network: Comput. Neu. Sys. 14,
103-118.

\item[-] Krahe R., Gabbiani F. (2004). Burst firing in sensory
systems. Nat. Rev. Neurosci. 5, 13--23.

\item[-] Latham P. E., and Niremberg S. (2005). Synergy,
Redundancy and Independence in Population Codes, Revisited. J.
Neurosci. 25, 5195--5206.

\item[-] Lesica N. A., and Stanley G. B. (2004). Encoding of
Natural Scene Movies by Tonic and Burst Spikes in the Lateral
Geniculate Nucleus. J. Neurosci. 24, 10731-10740.

\item[-] Martinez-Conde S., Macknik S. L., Hubel D. H. (2002). The
function of bursts of spikes during visual fixation in the awake
primate lateral geniculate nucleus and primary visual cortex.
Proc. Nac. Acad. Sci. 99, 13920–-13925.

\item[-] Meinen, Z. F., and Sejnowski, T. J. (1995). Reliability
of spike timing in cortical neurons. Science 268, 1503--1506.

\item[-] Metzner W., Koch C., Wessel R., and Gabbiani F. (1998).
Feature Extraction by Burst-Like Spike Patterns in Multiple
Sensory Maps. J. Neurosci. 18, 2283-2300.

\item[-] Montemurro, M. A., Panzeri, S., Maravall, M., Alenda, A.,
Bale, M. R., Bramilla, M., and Petersen, R. S. (2007a). Role of
precise spike timing in coding of dynamic vibrissa stimuli in
somatosensory thalamus. J. Neurophysiol. 98, 1871--1882.

\item[-] Montemurro, M. A., Senatore, R., and Panzeri, S. (2007b).
Tight Data-Robust Bounds to Mutual Information Combining Shuffling
and Model Selection Techniques. Neural Comput. 19, 2913--2957.

\item[-] Oertle, D. (1999). The role of timing in the brain stem
of vertebrate auditory nuclei. Annu. Rev. Physiol. 61, 497–519.

\item[-] Oswald A. M. M., Chacron M. J., Doiron B., Bastian J.,
and Maler L. (2004). Parallel Processing of Sensory Input by
Bursts and Isolated Spikes. J. Neurosci. 24, 4351-4362.

\item[-] Reich, D. S., Mechler, F., Purpura, K. P., and Victor, J.
D. (2000). Interspike Intervals, Receptive Fields, and Information
Encoding in Primary Visual Cortex. J. Neurosci. 20, 1964–-1974.

\item[-] Reinagel P., Godwin D., Sherman S. M., Koch C. (1999).
Encoding of Visual Information by LGN Bursts. J. Neurophysiol. 81,
2558-2569.

\item[-] Reinagel P., and Reid R. C. (2000). Temporal coding of
visual information in the thalamus. J. Neurosci. 20, 5392-5400.

\item[-] Rieke F., Warland D., de Ruyter van Steveninck R., Bialek
W. (1997). Spikes: Exploring the Neural Code. (Cambridge: MIT
Press).

\item[-] Rokem, A., Watzl, S., Gollisch, T., Stemmler, M., Herz,
A. V. M., and Samengo, I. (2006). Spike-Timing Precision Underlies
the Coding Efficiency of Auditory Receptor Neurons. J.
Neurophysiol. 95, 2541--2552.

\item[-] Sherman S. M. (2001). Tonic and burst firing: dual modes
of thalamocortical relay. Trends. Neurosci. 24, 122--126.

\item[-]  Schneidman, E., Bialek, W., and Berry, M. J. (2003).
Synergy, Redundancy, and Independence in Population Codes. J.
Neurosci. 23, 11539-11553.

\item[-] Strong, S. P., Koberle, R., de Ruyter van Steveninck, R.
R., and Bialek, W. (1998). Entropy and Information in Neural Spike
Trains. Phys. Rev. Lett. 80, 197--200.

\item[-] Theunissen, F., and Miller, J. P. (1995). Temporal
encoding in nervous systems: a rigorous definition. J. Comput.
Neurosci. 2, 149–162.

\end{enumerate}

\pagebreak

\section*{Key Concepts}

{\bf Firing-rate codes}: neural codes where all the information is
encoded in the time-dependent firing rate, additional response
properties being irrelevant.

\noindent {\bf Inter-spike interval (ISI)}: the interval between
two contiguous action potentials.

\noindent {\bf Pattern-based codes}: neural codes structured into
patterns of spikes, defined by characteristic correlations in
their relative timing.

\noindent {\bf Burst} of spikes: sequence of action potentials
separated by small inter-spike intervals, near to the refractory
period of the neuron.

\noindent {\bf Burst-triggered average (BTA)} of a burst of $n$
spikes: the average of stimulus stretches preceding the generation
of bursts of exactly $n$ spikes.

\noindent {\bf Mutual information} between stimuli and neural
responses: the amount of knowledge that can be gained about the
stimulus by observing the neural activity (and vice versa).

\end{document}